\documentclass[pra,aps,superscriptaddress,reprint]{revtex4-1}
\usepackage{amsmath, amsfonts, amssymb, mathrsfs}
\usepackage{graphicx,color,fancybox}
\usepackage{soul} 

\begin{document}

\title{Prediction of interface states in liquid surface waves with one-dimensional modulation}

\author{Xi Shi}
\email[Corresponding author:~]{xishi@shnu.edu.cn}
\affiliation{Department of Physics, Shanghai Normal University, Shanghai 200234, China}
\affiliation{MOE Key Laboratory of Advanced Micro-Structured Materials, School of Physics Science and Engineering, Tongji University, Shanghai 200092, China}

\author{Yong Sun}
\affiliation{MOE Key Laboratory of Advanced Micro-Structured Materials, School of Physics Science and Engineering, Tongji University, Shanghai 200092, China}

\author{Chunhua Xue}
\affiliation{MOE Key Laboratory of Advanced Micro-Structured Materials, School of Physics Science and Engineering, Tongji University, Shanghai 200092, China}

\author{Xinhua Hu}
\email[Corresponding author:~]{huxh@fudan.edu.cn}
\affiliation{Department of Materials Science and Key Laboratory of Micro- and Nano-Photonic Structures (Ministry of Education), Fudan University, Shanghai 200433, China}

\begin{abstract}
We theoretically studied the interface states of liquid surface waves propagating through the heterojunctions formed by a bottom with one-dimensional periodic undulations. Via considering the periodic structure as a homogeneous one, our systematic study shows that the signs of the effective depth and gravitational acceleration are opposite within the band gaps no matter the structure is symmetric or asymmetric. Those effective parameters can be used to predict the interface states which could amplify the amplitudes of liquid surface waves. These phenomena provide new opportunities to control the localization of water-wave energy.
\end{abstract}

\maketitle

\section{Introduction}

In recent decades the propagation of liquid surface waves (LSWs) over an uneven bottom with a periodic modulation, such as rippled bottoms, periodic drilled holes as well as periodic arrays of surface scatters, has attracted a great deal of attentions. Originating from the Bragg resonances of water waves, the periodic structure with the scale of the half-wavelength can strongly reflect the water waves~\cite{hu2003complete,hu2003band}, and thus exhibited many peculiar phenomena including water wave blocking~\cite{jeong2004experimental,shen2005observation,tang2006omnidirectional}, superlensing effects~\cite{hu2004superlensing}, self-collimation~\cite{shen2005self} and directional radiation~\cite{mei2009highly,mei2010enhanced,wang2013experimental}. Recently, a concept of effective liquid was developed for thoroughly understanding the interaction of water waves with periodic structures~\cite{hu2005refraction,hu2011negative}. Theoretical studies have shown that the effective gravity in water pierced by a cylinder array is larger than the one on the  earth, which induces a new type of water wave refraction~\cite{hu2005refraction}.  For water with a resonator array, the effective gravity $g_e$ can be negative near resonant frequency, so that water waves cannot propagate through the array~\cite{hu2011negative,hu2013experimental}. More particularly, the effective gravity $g_e$ even can be infinite when water is covered by a thick, rigid and unmovable plate, which can be used for broadband focusing and collimation of water waves~\cite{zhang2014broadband}. These results promise a new mechanism to control the propagation of water waves, which exhibits particular applications in wave energy conversion and coastal protection~\cite{budar1975resonant,callaway2007energy,cruz2007ocean,engstrom2009wave,garnaud2009bragg}.

Most studies concerning liquid waves have, hitherto, been focused on the effective gravity of water in period structures. In fact, the water-wave process is affected not only the gravity of the earth but also the depth of water. Therefore, the effective theory for liquid waves includes not only the effective gravity $g_e$ but also the effective depth $h_e$. However, the latter has rarely been paid much attention~\cite{hu2005refraction,hu2011negative}. Here we further explore the interactions of water waves with periodic structures by utilizing the concept of the effective gravity $g_e$ and effective depth $h_e$. Our results show that the band gaps of LSW propagating over rippled bottom can be characterized with either a negative depth with a positive gravitational acceleration (negative depth bottom, NDB), or a negative gravitational acceleration with a positive depth (negative gravity bottom, NGB). In the NDB-NGB pair structure, the interface states of LSW are realized accurately under the matching conditions with respect to the effective impedance and effective phase shift. Theoretical calculations show that the LSW are strongly localized at the interface, with one order of magnitude enhancement, which possesses potential applications in wave energy conversion. Recent results about classical wave periodic systems show that materials with two different single negative parameters (for example, electromagnetic materials with negative permittivity or negative permeability) have different topological orders~\cite{mei1989theory,tan2014photonic,xiao2014surface,xiao2015geometric,shi2016topological,yang2016topological}. Hence, interface states formed at the boundary separating two periodic structures having different band gap topological characteristics. Our results may provide another insight to understand the band gap properties in LSWs.

The paper is organized as follows. In Sec. \ref{sec:effective}, we present the method used for the retrieval of effective depth and effective gravitational acceleration for LSWs. The effective parameters for LSWs within band gaps are discussed with the retrieval method. In Sec. III, we explore the interface state existing at the boundary between two periodic bottoms with different effective parameters. Here two types of paired structures are discussed: one is composed of asymmetric unit cells, and the other is symmetric unit cells. In Sec. IV, we investigate the field enhancement behavior for the interface state. Finally, a conclusion is given in Sec. V.

\section{Effective parameters for liquid surface wave in band gaps~\label{sec:effective}}

\begin{figure*}[t]
\centering
\includegraphics[scale=1]{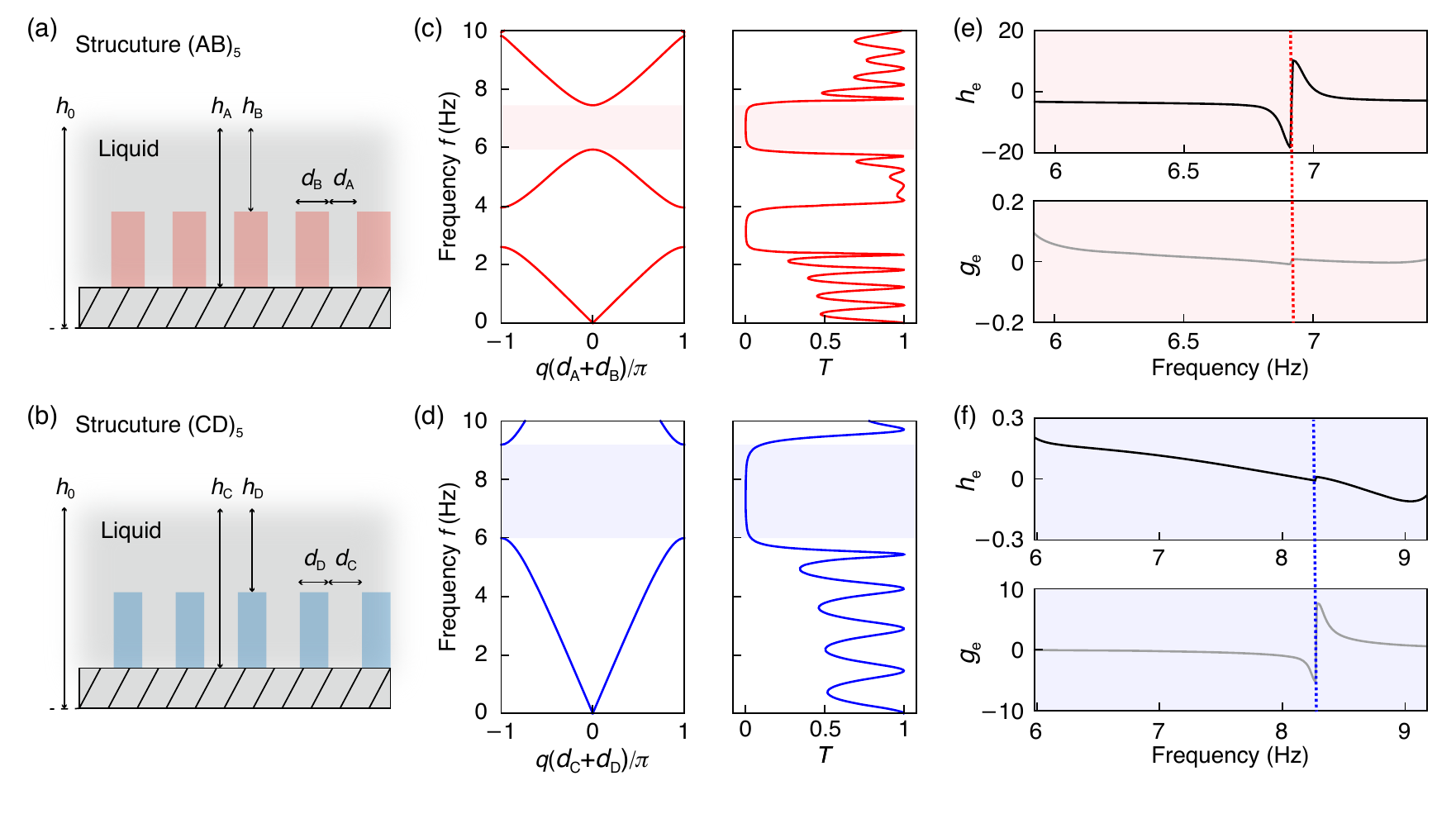}\\
\caption
{Schematic views of (a) the finite periodic rippled bottoms (AB)$_5$, (b) the finite periodic rippled bottoms (CD)$_5$. (c), (d) Band structures and transmission spectra of  (AB)$_5$ and  (CD)$_5$. The retrieved effective gravitational acceleration $g_e$  and effective depth  $h_e$ of  (AB)$_5$ and (CD)$_5$ within band gaps are shown in (e) and (f).}
\label{fig:1}
\end{figure*}

In this section, we will explore the effective parameters of LSW system using the retrieval method. This concept has been introduced firstly by Hu \textit{et al}~\cite{hu2005refraction}. For the linear, inviscid, irrotational and shallow LSW over an uneven bottom, the wave equation satisfies~\cite{mei1989theory,dingemans1997water,ye2003water} (rigorous when $kh \ll 1$),
\begin{equation}
\left( \nabla \cdot h \nabla + \frac{\omega^2}{g} \right) \eta
= 0,
\end{equation}
where $\eta$ is the displacement of the liquid surface, $\omega$ is the angular frequency, $g$ is the gravitational acceleration, and $h$ is the liquid depth. If the LSW system is taken as an even bottom with same reflection and transmission coefficients, it has the same effective gravitational acceleration and depth as the artificial homogeneous liquid. Supposing that the plane LSW normal incident into a one dimensional (1D) structure along the $x$ direction, and thus the surface displacement of LSW $\eta$ can be written as the superposition of forward and backward waves $\eta = A e^{ikx} + B e^{-ikx}$, with $A$ and $B$ the amplitudes of the forward and backward waves, respectively. Supposing that the properties of the structure can be described with the effective parameters, namely the effective relative depth $h_e$ and effective relative gravitational acceleration $g_e$. The effective index and the effective impedance can be written as $n_e = 1 / \sqrt{g_e h_e}$ and $z_e = \sqrt{h_e / g_e}$, respectively~\cite{hu2005refraction}. For the hypothetic artificial homogeneous liquid, the reflection coefficient $r$ and transmission coefficient $t$ can be obtained by using transfer matrix method~\cite{zhang2012band}
\begin{subequations}
\begin{align}
\frac{1}{t}
&= \cos(n_e k_0 d) - \frac{i}{2} \left( z_e + \frac{1}{z_e} \right) \sin(n_e k_0 d), \\
\frac{r}{t}
&= - \frac{i}{2} \left( z_e - \frac{1}{z_e} \right) \sin(n_e k_0 d),
\end{align}
\end{subequations}
where $k_0$ is the wave vector within background liquid depth, $d$ is the total thickness of structure. Through a standard retrieval procedure the effective index $n_e$ and effective impedance of water wave $z_e$ are given as follows
\begin{equation}\label{eq:effective_impedance}
z_e =
\pm \sqrt{\frac{(1 + r)^2 - t^2}{(1 - r)^2 - t^2}}
\end{equation}
and
\begin{equation}
e^{i n_e k_0 d} =
\frac{(z_e + 1) t}{z_e + 1 - (z_e - 1) t},
\end{equation}
where $n_e = \frac{1}{k_0 d} \{ \text{Im}[\ln (e^{i n_e k_0 d})] + 2 m \pi - i \text{Re}[\ln (e^{i n_e k_0 d})] \}$ with $m$ an integer. The sign on the right-hand side in Eq. \eqref{eq:effective_impedance} can be determined by $\text{Re} (z_e) \geq 0$, $\text{Im} (z_e) \leq 0$. Then the effective parameters $h_e$ and $g_e$ can be obtained by $h_e = z_e / n_e$ and $g_e = 1 / (z_e n_e)$. In this way, the 1D structure can act as an even bottom with the effective relative depth $h_e$ and the effective relative gravitational acceleration $g_e$.

It is known that the periodic modulation of liquid depth over the bottom would lead to the band structure of LSW, and the wave propagation is forbidden in the band gap~\cite{tang2006omnidirectional}. In the following, we use the effective parameters to investigate the band gap of 1D periodic structure. Two different 1D periodic structures, (AB)$_5$ and (CD)$_5$, are studied, as shown in Fig. \ref{fig:1} (a) and (b), respectively. Here A and C represent the valleys with the width of $d_A = 9.2$ mm and $d_C = 4.9$ mm, while B and D represent the ridges the width of $d_B = 11$ mm and $d_D = 4.2$ mm. Subscript 5 is the periodic number. The liquid depth over the valleys and ridges are $h_A = h_C = 6$ mm and $h_B = h_D = 1$ mm, respectively. The liquid depth of the background is $h_0 = 10$ mm. The liquid is chosen as CFC-113 of the Dupont company, a popular solvent with surface tension 17.3 dyn/cm and density 1.48 g/cm$^3$. We use this liquid instead of water since this liquid has a very low capillary length and small dissipation. Thus the phenomena of LSW can easily be observed in experiments~\cite{hu2004superlensing,shen2005self}.

\begin{figure}[b]
\centering
\includegraphics[scale=1]{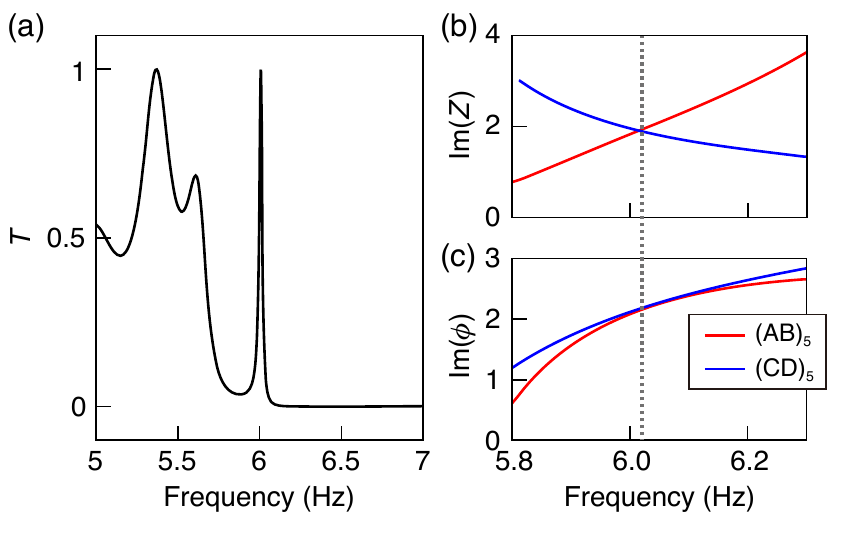}\\
\caption
{(a) Transmission spectrum of the heterostructure (AB)$_5$(CD)$_5$. (b) Imaginary impedance of (AB)$_5$ and (CD)$_5$ (the sign of the imaginary parts of (CD)$_5$ have been reversed). (c)  Imaginary phase of (AB)$_5$ and (CD)$_5$. Matching conditions are satisfied at 6.01 Hz, as indicated by the vertical gray dashed line.}
{\label{fig:2}}
\end{figure}

The band strcture and  transmission of (AB)$_5$ are shown in Fig.\ref{fig:1}(c). In this figure, there are two band gaps in this structure from low to high frequency region. The band gap  from 5.92 to 7.45 Hz is our concerned region and retrieved effective parameters are given in Fig.\ref{fig:1} (e). (AB)$_5$ has effective single negative parameters in the band gap, for the wave incident from right side. The effective parameters satisfy $h_e < 0$ and $g_e > 0$ in the region from 5.92 to 6.91 Hz, while $h_e > 0$ and $g_e < 0$ in the region from 6.91 to 7.45 Hz. The similar calculations about (CD)$_5$  is given in Fig. \ref{fig:1}(d) and (f). Clearly there is a band gap from 5.98 to 9.18 Hz. Using the retrieval method, the effective liquid depth $h_e$ and gravitational constant $g_e$ of (CD)$_5$ is plotted. It shows that (CD)$_5$ have effective single negative parameters in the band gap, for the wave incident from left side. In details, $h_e > 0$ and $g_e < 0$ in the region from 5.98 to 8.30 Hz, while $h_e < 0$ and $g_e > 0$ in the region from 8.30 to 9.18 Hz. Thus far, the results in (AB)$_5$ and (CD)$_5$ show different behavior from lower frequency to higher in band gaps. For (AB)$_5$ , it acts as a NDB at lower frequency and a NGB at higher frequency. For (CD)$_5$, it shows opposite behavior in band gap. The band gap of finite periodic rippled bottoms could be characterized by the effective negative depth and the effective negative gravitational acceleration with parameters $g_e h_e < 0$.

\section{Interface states in liquid surface wave}

The results above show that the LSW period system in band gap can prevent the propagation of water wave and may mimic two types of effective structure with parameters $h_e < 0$, $g_e > 0$ and $h_e > 0$, $g_e < 0$ respectively. In the following, we investigate the interface state existing in the pair of NDB and NGB structure.

Using the transfer matrix method, we can obtain the matching condition between NDB and NGB structures as
\begin{subequations}
\begin{align}
\text{Im} (z_\text{NDB})
&= - \text{Im} (z_\text{NGB}), \\
\text{Im} (n_\text{NDB} k_0 d_\text{NDB})
&= \text{Im} (n_\text{NGB} k_0 d_\text{NGB}),
\end{align}
\end{subequations}
where $z$ is the effective impedance of LSW retrieved from transmission and reflection coefficients of corresponding structures, and $n$ and $d$ denote the refractive index and thickness, respectively; Im represents the imaginary part. Therefore, for the NDB-NGB pair structure, the perfect transmission of LSW can occur at the frequency which meets the matching condition.

\begin{figure}[t]
\centering
\includegraphics[scale=1]{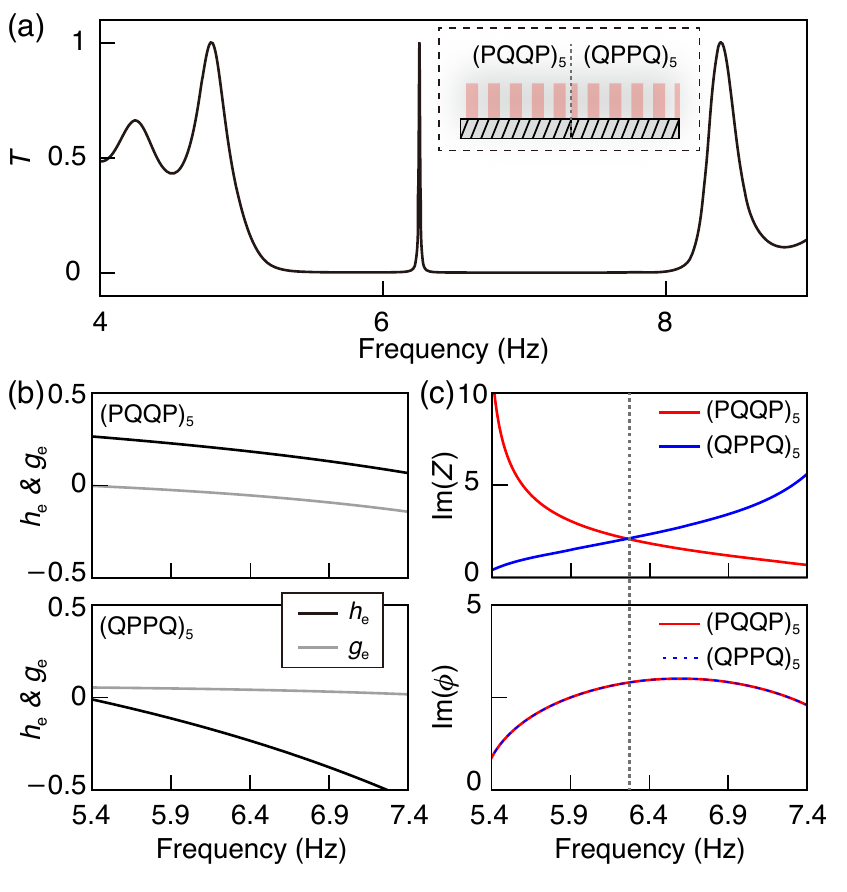}\\
\caption
{(a) Transmission spectrum of the heterostructure (PQQP)$_5$(QPPQ)$_5$. A schematic view of the structure is shown in the inset. (b) The retrieved effective parameters  $g_e$ and $h_e$  versus frequency for (PQQP)$_5$ and (QPPQ)$_5$. (c) Imaginary impedance of(PQQP)$_5$ and (QPPQ)$_5$ for upper panel (the sign of the imaginary parts of (QPPQ)$_5$  have been reversed). Imaginary phase of (PQQP)$_5$ and (QPPQ)$_5$ for lower panel. Matching conditions are satisfied at 6.26 Hz, as indicated by the vertical dashed line. }
{\label{fig:3}}
\end{figure}

To explore the perfect transmission behavior of LSW under the matching condition, we give the imaginary parts of the effective impedances and the effective phase shifts of (AB)$_5$ and (CD)$_5$ in Fig. \ref{fig:2}(b) and (c). As indicated by the vertical dashed line, the matching conditions are satisfied at the frequency of $f = 6.01$ Hz, where (AB)$_5$ acts as a NDB structure and (CD)$_5$ acts as a NGB. Thus one can infer that there would be an interface state of LSW at 6.01 Hz for the paired structure (AB)$_5$(CD)$_5$. The transmission of (AB)$_5$(CD)$_5$ is given in Fig \ref{fig:2}(a).A perfect transmission peak emerges around $f = 6.01$ Hz, which is original in the band gap of (AB)$_5$ and (CD)$_5$. It means the LSW tunnels through the heterostructure although it cannot propagate through (AB)$_5$ or (CD)$_5$.

The discussion above is about the periodic modulated bottoms composed of asymmetric unit cells. Next, we will introduce a special type of structure composed of symmetric unit cells, i.e. the structures (PQQP)$_5$, (QPPQ)$_5$ and the heterostructure (PQQP)$_5$(QPPQ)$_5$, as is shown in figure \ref{fig:3}(a). Here P and Q correspond to valley and ridge with the same width 2.5 mm. The liquid depth over valley and ridge are set to be 4 mm and 1 mm, respectively. The transmission spectra of the structure (PQQP)$_5$ and (QPPQ)$_5$, as is illustrated in figure 6(b), exhibit the same region of band gap from 5.6 to 8.4 Hz. Moreover, it can be seen from Fig. \ref{fig:3}(b) that the effective parameters of (PQQP)$_5$ satisfy $h_e > 0$ and $g_e < 0$ throughout the whole band gap. Thus the structure (PQQP)$_5$ can mimic a NGB structure. On the contrary, (PQQP)$_5$ can mimic a NDB structure throughout its whole band gap, i.e. $h_e < 0$ and $g_e > 0$. It has been mentioned, for the 1D periodic structures composed of asymmetric unit cells, the band gap is divide into two regimes with respect to the retrieved effective parameters, which corresponds to NGB and NDB, respectively. Here, for 1D periodic structures composed of symmetric unit cells, the retrieved effective parameters throughout the whole band gap would only be one type, i.e. $h_e > 0$ and $g_e < 0$ for (PQQP)$_5$, and $h_e <0$ and $g_e > 0$ for (QPPQ)$_5$. Since the structure (PQQP)$_5$ and (QPPQ)$_5$ respectively corresponds to NGB and NDB, we can expect the realization of the imaginary impedance matching and imaginary phase matching between these two structures. The imaginary impedance and imaginary phase of (PQQP)$_5$  and (QPPQ)$_5$  are calculated in figure \ref{fig:3}(c). On the one hand, these two structures have the same imaginary phase in band gap, which indicates the automatic satisfaction of the imaginary phase matching. Consequently, the appearance of tunneling effect is only determined by the imaginary impedance matching, which means that the interface states could be obtained more easily with symmetric unit cells. On the other hand, the imaginary impedance matching is satisfied at 6.26 Hz, which indicates the tunneling frequency.  As a result, for the heterostructure (PQQP)$_5$(QPPQ)$_5$, an interface state with high transmission can be observed at 6.26 Hz, which locates in the first gaps of (PQQP)$_5$ and (PQQP)$_5$, as is shown in figure \ref{fig:3}(a).

\section{Field distribution of transmission spectra}
\begin{figure}[b]
\centering
\includegraphics[scale=1]{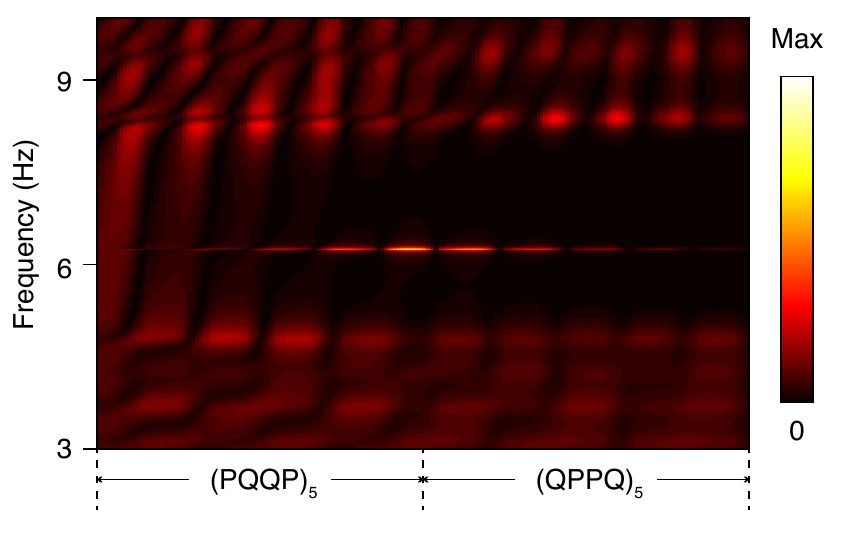}\\
\caption
{The field intensity of (PQQP)$_5$(QPPQ)$_5$. Horizontal axle is the length of structure. The amplitude of LSW is enhanced about 25 times near the interface. }
{\label{fig:4}}
\end{figure}
In this section we will investigate the field enhancement of (PQQP)$_5$(QPPQ)$_5$ induced by the interface state. In our calculations, the amplitude of incident LSW is normalized. In figure 8(a), the field distributions of (PQQP)$_5$(QPPQ)$_5$ are given for the frequencies from 3 to 10 Hz. It shows that the LSW cannot propagate through the structure in the  frequency region of band gaps (5 Hz to 8.9 Hz). However, the fields show that LSW can clearly tunnel through the structure at 6.26 Hz, which corresponds to the frequency of interface state. When the interface state is excited, LSW is primarily localized at the interface and exponentially decays from the interface to both ends, which is in detail illustrated in figure \ref{fig:3}. Our results show that the amplitude of LSW at the interface between (PQQP)$_5$ and (QPPQ)$_5$ is effectively driven up as large as 25 times than that of incident wave.

\section{Conclusion}

In summary, we theoretically studied the propagation of LSWs over a bottom with a one-dimensional periodic undulation. The results reveal that the signs of the effective depth and gravitational acceleration are opposite within the band gaps. Under the conditions of the impedance matching and phase matching, an interface states can be realized at the interface between the structure with negative $h_{e}$ and the structure with negative $g_{e}$. Moreover, the LSW energy localizations at the interface can be enhanced over one order of magnitude than that of the incident LSW. Our work possesses potential applications in the utilization of energy in water wave. In addition, these results may pave the way for realizing many exotic phenomena based on single negative materials and zero-refractive-index materials.

\begin{acknowledgments}
The authors are grateful to Ang Chen for helpful discussions. This work is supported by  the NSFC (No. 61422504, 11474221, 11234010 , 11204217 and 11704254) and financial support from the China Scholarship Council (Grant No. 201706265021).
\end{acknowledgments}


%

\end{document}